\documentclass[aps,showpacs,prep,graphics,twocolumn]{revtex4}
\usepackage{amssymb}
\usepackage[dvips]{graphicx}
\usepackage{amsmath}
\usepackage{bm}
\usepackage{epsfig}

\begin{document}

\title{ Scalar field unification of interacting viscous dark fluid from a geometrical scalar-tensor theory of gravity.}

\author{ $^{1}$ Jos\'e Edgar Madriz Aguilar\thanks{E-mail address: madriz@mdp.edu.ar}, $^{2}$ A. Gil-Ocaranza,  $^{1}$ M. Montes and $^{2}$ J. Zamarripa
\thanks{E-mail address: mariana.montnav@gmail.com} }
\affiliation{$^{1}$ Departamento de Matem\'aticas, Centro Universitario de Ciencias Exactas e ingenier\'{i}as (CUCEI),
Universidad de Guadalajara (UdG), Av. Revoluci\'on 1500 S.R. 44430, Guadalajara, Jalisco, M\'exico.  \\
and\\
$^{2}$ Centro Universitario de los Valles\\
Carretera Guadalajara-Ameca Km. 45.5, C.P. 46600, Ameca, Jalisco, M\'exico.\\
E-mail:  jose.madriz@academicos.udg.mx, 
madriz@mdp.edu.ar,
 mariana.montes@academicos.udg.mx}

\begin{abstract}
	
 We investigate, in the framework of a recently introduced new class of invariant geometrical scalar-tensor theory of gravity, the possibility that a viscous dark fluid can be described in a unified manner by a single scalar field. Thus we developed a model in which both the metric tensor and the scalar field have geometrical origin. The scalar field is characterized by a non-canonical kinetic term and the scalar viscosity of the dark fluid appears as soon the kinetic energy of the scalar field is no longer canonical. The scalar viscosity is considered as a function of the Hubble and the deceleration parameters. To illustrate the formalism we have considered two cases: a constant and  a thermodynamic equation of state parameters. In the both cases we obtain analytic representations for the scalar field and their respective potentials. We delimit free parameters by comparing with some Planck 2018 results.

\end{abstract}

\maketitle

\vskip .5cm
 Weyl-Integrable geometry, scalar-tensor gravity, viscous interacting dark fluid, accelerated cosmic expansion, dark energy and dark matter.

\section{Introduction}

An explanation for the acceleration in the cosmic expansión continues being a challenge in modern cosmology \cite{rea1}. Such acceleration has been corroborated by Ia Supernovae data \cite{rea2,rea3}, baryonic acoustic oscillations (BAO) \cite{rea4} and Cosmic Microwave Background (CMB) anisotropies \cite{rea5,rea6,rea7}. In the quest for an explanation of the origin of such acceleration the main proposals are divided in modified theories of gravity and dark energy models \cite{rea8,rea9} . In the second branch we can find models in which he dark energy is considered as a fluid with viscosity where thermodynamic effects are also important \cite{rea1,rea10,rea11,rea12,rea13,rea14,rea15,rea16,rea17,rea18}. Dark energy models treated as imperfect fluids can be considered more realistic \cite{rea19}. In fact, one characteristic of dark energy models with perfect fluids is that as the dark energy component has negative pressure, the matter component has null pressure and the radiation pressure is $\rho_r/3$, the total pressure is negative and in this sense the material content of the universe violate the strong energy condition \cite{rea20}. In addition, some observational data suggest that the dark energy equation of state could be time varying, and thus a perfect fluid prescription can suffer of some thermodynamic problems linked to the positiveness of the entropy and temperature \cite{rea20,rea21,rea22}. Viscous dark energy models can avoid that kind of problems. \\

However, parallel to dark energy models, modified theories of gravity have been a recourse to explain the present cosmological scenario of accelerating expansion, as for example, scalar-tensor theories of gravity \cite{rea23,rea24}, $f(R)$ theories \cite{rea25,rea26} and theories with extra dimensions \cite{rea27,rea28}, among others. Recently a new approach of scalar-tensor theories of gravity has been proposed. This new approach is known as geometrical scalar-tensor theories of gravity \cite{rea29,rea30}. One of the main motivations for these theories is to avoid the controversy on which of the Einstein or Jordan frames is the physical one. The controversy is related to the fact that the way of passing from one frame to the other changes the background geometry and makes that geodesic obervers in one frame are not in the other.  In that approach the symmetries of the action and of the background geometry i.e. the compatibility  condition, are the same, and thus the scalar field becomes part of the affine structure of the space-time manifold. In this sense, both the metric and the scalar field are geometric in nature and hence the controversy can be alleviated \cite{rea29,rea30,rea31}. In geometrical scalar-tensor theories the background geometry is not fixed apriori instead it is determined by the Palatini's principle \cite{rea29,rea30}. Different applications of these theories have been done. For example, inflationary cosmology \cite{rea32,rea33}, quintessence, cosmic magnetic fields, and some cosmological models have been studied topics of these theories \cite{rea34,rea35,rea36}.\\

In this letter our interest is to derive a viscous dark fluid cosmological model in the setting of geometrical scalar-tensor theories of gravity, which can be described in a unified manner by a single scalar field of geometrical origin. In this derivation we will assume that both dark sectors: dark matter and dark energy can interact each other.
In section I we give a brief introduction. Section II is devoted to the formalism of geometrical scalar-tensor theories. In section III we obtain the field equations when matter sources are present. In section IV we develope a viscous dark fluid model. We left section V to give some final remarks.\\

\section {The formalism of geometrical scalar-tensor gravity }

 We consider a scalar-tensor theory of gravity in vacuum  given by the action
\begin{equation}\label{f1}
S=\frac{1}{16\pi}\int d^{4}x\sqrt{-g}\left\lbrace \Phi {\cal R}+\frac{\tilde{\omega}(\Phi)}{\Phi}g^{\mu\nu}\Phi_{,\mu}\Phi_{,\nu}-\tilde{V}(\Phi)\right\rbrace ,
\end{equation}
where ${\cal R}$ denotes the Ricci scalar, $\tilde{\omega}(\Phi)$ is a function of the scalar field $\Phi$ and $\tilde{V}(\Phi)$ is a scalar potential. By means of the field transformation  $\varphi=-\ln (G\Phi)$, the action (\ref{f1}) can be rewritten as \cite{rea32}
\begin{equation}\label{f2}
S=\int d^{4}x\sqrt{-g}\left\lbrace e^{-\varphi}\left[\frac{{\cal R}}{16\pi G}+\frac{1}{2}\omega (\varphi)g^{\mu\nu}\varphi_{,\mu}\varphi_{,\nu}\right]-V(\varphi)\right\rbrace,
\end{equation}
with the identifications $(1/2)\omega(\varphi)=(16\pi G)^{-1}\tilde{\omega}[\varphi(\Phi)]$ and $V(\varphi)=(16\pi)^{-1}\tilde{V}(\varphi(\Phi))$.  Varying the action \eqref{f2}  with respect to the affine connection following a Palatini variational principle, it yields  \cite{rea29}
\begin{equation}\label{f3}
\nabla _{\mu}g_{\alpha\beta}=\varphi_{,\mu}g_{\alpha\beta}.
\end{equation}
This compatibility condition indicates that the background geometry corresponding to the action \eqref{f2} is the Weyl-integrable one. Thus to distinguish  this covariant derivative from the Riemannian one, we will denote it as $^{(w)}\nabla$. Applying, at the same time, the transformations
\begin{eqnarray}\label{f4}
\bar{g}_{\alpha\beta}&=&e^{f}g_{\alpha\beta}\\
\label{f5}
\bar{\varphi}&=&\varphi +f,
\end{eqnarray}
where $f=f(x^{\alpha})$ is a well defined function of the spacetime coordinates, the condition \eqref{f3} results to be invariant. As the Weyl-Integrable geometry is the background geometry, then the action \eqref{f2} must be invariant under \eqref{f4} and \eqref{f5}. However, it is not difficult to verify that \eqref{f2} is not.  Thus, we introduce the invariant action
\begin{small}
	\begin{eqnarray}
	{\cal S}&=&\int d^{4}x\sqrt{-g}\,e^{-\varphi}\left[\frac{{\cal R}}{16\pi G}+\frac{1}{2}\omega(\varphi)g^{\alpha\beta}\varphi_{:\alpha}\varphi_{:\beta}-\right.\nonumber\\
	&&
	\left.V(\varphi)e^{-\varphi}-\frac{1}{4}H_{\alpha\beta}H^{\alpha\beta}e^{-\varphi}\right],\label{f9}
	\end{eqnarray}
\end{small}
where we have introduce the gauge covariant derivative
$\varphi_{:\mu}=(\,^{(w)}\nabla_{\mu}+\gamma B_{\mu})\varphi$,
with $B_{\mu}$ being a gauge vector field and  $\gamma$ is a purely imaginary coupling constant. The $H_{\alpha\beta}=W_{\beta ,\alpha}-W_{\alpha ,\beta}$ is denoting the field strength of the gauge boson field $W_{\mu}=\varphi B_{\mu}$.  The invariance under (\ref{f4}) and (\ref{f5}) of (\ref{f9}) requires the following transformation rules to hold
\begin{eqnarray}\label{f10a}
\bar{\varphi}\bar{B}_{\mu} &=& \varphi B_{\mu}-\gamma^{-1}f_{,\mu},\\
\bar{\omega}(\bar{\varphi})&=&\omega(\bar{\varphi}-f)=\omega(\varphi),\label{f10b}\\
\bar{V}(\varphi)&=& V(\bar{\varphi}-f)=V(\varphi).\label{f10c}
\end{eqnarray}
On the other hand, all the previous equations have been formulated on the Weyl frame. We mean by Weyl frame the set $(M,g,\varphi, B_{\alpha})$, where the background geometry is the corresponding to \eqref{f3}. Thus, the transformations (\ref{f4}) and (\ref{f5}) allow to pass from one frame to another, both Weylian. However, in the particular frame where $f=-\varphi$, we can write the condition \eqref{f3} as $\nabla_{\lambda} h_{\alpha\beta}=0$, which clearly corresponds to a Riemannian  geometry with respect to the effective metric $h_{\mu\nu}\equiv\bar{g}_{\mu\nu}=e^{-\varphi}g_{\mu\nu}$. This particular frame is known as the Riemann frame and is denoted by $(M,\bar{g},\bar{\varphi}=0,\bar{B}_{\alpha})=(M,h,A_{\alpha})$. We use this terminology to differenciate it from the traditional Jordan and Einstein frames in usual scalar-tensor theories, given that in the former geodesics are not invariant under conformal transformations, while in geometrical scalar-tensor theories geodesics are Weyl invariant \cite{rea32}.\\

In the Weyl frame both the scalar field $\varphi$ and the gauge vector field $B_\mu$ are geometrical because they form part of the affine structure of the space-time manifold. In the Riemann frame the scalar field $\phi(x)$ and the vector field $A_{\mu}$ are physical quantities. Notice that when we pass to the Riemann frame the field $\varphi$ is renamed as $\phi(x)$ just to emphasize that is the physical and not the geometrical scalar field. The same consideration is for the vector field: $B_{\alpha}\rightarrow A_{\alpha}$.\\

The action (\ref{f9}) written in the Riemann frame has the form
\begin{eqnarray}
{\cal S}&=&\int d^{4}x\sqrt{-h}\left[\frac{R}{16\pi G}+\frac{1}{2}\omega(\phi)h^{\alpha\beta}{\cal D}_{\alpha}\phi{\cal D}_{\beta}\phi-V(\phi)\right.\nonumber\\
\label{Rie2}
&&\left. -\frac{1}{4}F_{\alpha\beta}F^{\alpha\beta}\right],
\end{eqnarray}
where ${\cal D}_{\mu}=\nabla_{\mu}+\gamma A_{\mu}$ with the operator $\nabla_{\lambda}$ denoting the Riemannian covariant derivative and $F_{\mu\nu}=A_{\nu ,\mu}-A_{\mu,\nu}$. 
The last term in the action \eqref{Rie2} is invariant under the field transformation
\begin{equation}\label{yq1}
\overset{\smile}{A}_{\mu}=A_{\mu}-\gamma^{-1}\sigma_{,\mu},
\end{equation}
where $\sigma =\sigma (x^{\alpha})$. It suggests that $A_{\mu}$ can play the role of an electromagnetic potential. It is important to remember that the last term in \eqref{Rie2} comes from the last term in \eqref{f9}, which was introduced when the Weyl invariance of the action \eqref{f9} was imposed. \\

Now, to ensure the invariance under  (\ref{yq1}) of the complete action (\ref{Rie2}), the next internal symmetries must be valid
\begin{eqnarray}\label{yq2}
\overset{\smile}{\phi}&=&\phi e^{\sigma},\\
\label{yq2p}
\overset{\smile}{\omega}(\overset{\smile}{\phi})&\equiv& e^{-2\sigma}\omega(e^{-\sigma}\overset{\smile}{\phi})=\omega(\phi)\\
\label{yq2q}
\overset{\smile}{V}(\overset{\smile}{\phi})&\equiv& V(e^{-\sigma}\overset{\smile}{\phi})=V(\phi).
\end{eqnarray}
If we interpret $A_{\mu}$ as the electromagnetic potential, 
 the action (\ref{Rie2}) can be extended by adding a source term for $A_{\mu}$ in the form 
\begin{eqnarray}
{\cal S}&=&\int d^{4}x\sqrt{-h}\left[\frac{R}{16\pi G}+\frac{1}{2}\omega(\phi)h^{\alpha\beta}{\cal D}_{\alpha}\phi{\cal D}_{\beta}\phi-V(\phi)\right.\nonumber\\
&&
\left.-\frac{1}{4}F_{\alpha\beta}F^{\alpha\beta}-J^{\alpha}A_{\alpha}\right],\label{yq3}
\end{eqnarray}
where $J^{\mu}$ is a conserved current density. Thus, straithforward calculations show that the action (\ref{yq3}) leads to the field equations
\begin{eqnarray}
&& G_{\mu\nu}=-8\pi G \left[\omega(\phi){\cal D}_{\mu}\phi{\cal D}_{\nu}\phi-\frac{1}{2}h_{\mu\nu}\left(\omega(\phi)h^{\alpha\beta}{\cal D}_{\alpha}\phi{\cal D}_{\beta}\phi\right.\right.\nonumber\\
\label{Rie6}
&& \left.\left. - 2V(\phi)\right)-\tau_{\mu\nu}^{(em)}\right]\\
&& \omega(\phi)\Box\phi+\frac{1}{2}\omega^{\prime}(\phi)h^{\mu\nu}{\cal D}_{\mu}\phi{\cal D}_{\nu}\phi-\gamma\omega^{\prime}(\phi)A^{\mu}\phi{\cal D}_{\mu}\phi\nonumber\\
&&+\gamma\omega(\phi)\nabla_{\mu}A^{\mu}-\gamma^2\omega(\phi)A^{\mu}A_{\mu}\phi+V^{\prime}(\phi)=0,\label{Rie7}\\
&& \nabla_{\mu}F^{\mu\nu}= J^{\nu}-\gamma\omega(\phi)h^{\mu\nu}\phi{\cal D}_{\mu}\phi,\label{Rie8}
\end{eqnarray}
where  $\Box =h^{\mu\nu}\nabla_{\mu}\nabla_{\nu}$ denotes the usual D'Alambertian operator, $\tau_{\mu\nu}^{(em)}=T_{\mu\nu}^{(em)}-h_{\mu\nu}J^{\alpha}\! A_{\alpha}$, with $T_{\mu\nu}^{(em)}=F_{\nu\beta}F_{\mu}\,^{\beta}-\frac{1}{4}h_{\mu\nu}F_{\alpha\beta}F^{\alpha\beta}$ being the energy-momentum tensor for a free electromagnetic field. With the idea to study cosmological applications to this formalism in the next section we show the manner to introduce external sources of matter in the formalism. \\

\section{The field equations in presence of matter}

As it was shown in \cite{rea29}, a Weyl invariant action for matter sources can be written as
\begin{equation}\label{wm1}
S_{m}=\int d^{4}x\sqrt{-g}\,e^{-2\varphi}L_{m}\left(e^{-\varphi}g_{\mu\nu},\Psi,^{(w)}\!\nabla\Psi\right),
\end{equation}
with $\Psi$ denoting some matter field and  $L_m$ representing the matter lagrangian which is constructed taking into account the prescription $L_{m}(g,\varphi,\Psi,^{(w)}\!\nabla\Psi)\equiv L_{m}^{(sr)}(e^{-\varphi}g,\Psi,^{(w)}\!\nabla\Psi)$, where $L_{m}^{(sr)}$ stands for the lagrangian of $\Psi$ in the  Minkowski space-time. Hence , the energy-momentum tensor $T_{\mu\nu}(\varphi,g,\Psi, ^{(w)}\!\nabla\Psi)$ for matter sources in an arbitrary Weyl frame $(M,g,\varphi)$ is determined by
\begin{eqnarray}
\delta\int d^{4}x \sqrt{-g} e^{-2\varphi}L_{m}(\varphi,g_{\mu\nu}, \Psi,^{(w)}\!\nabla\Psi) =\nonumber \\
\label{wm2} \int d^{4}x \sqrt{-g}e^{-2\varphi}T_{\mu\nu}(\varphi,g_{\mu\nu},\Psi,^{(w)}\!\nabla\Psi)\delta(e^{\varphi} g^{\mu\nu}),
\end{eqnarray}
where $\delta$ denotes the variation with respect to both $g_{\mu\nu}$ and $\varphi$. Therefore, the field equations in the Riemann frame in presence of matter sources read
\begin{eqnarray}
&& G_{\mu\nu}=-8\pi G T_{\mu\nu}-8\pi G \left[\omega(\phi){\cal D}_{\mu}\phi{\cal D}_{\nu}\phi-\right. \nonumber \\
&&\left.\frac{1}{2}h_{\mu\nu}\left(\omega(\phi)h^{\alpha\beta}{\cal D}_{\alpha}\phi{\cal D}_{\beta}\phi- 2V(\phi)\right)-\tau_{\mu\nu}^{(em)}\right] \label{wm3}\\
&& \omega(\phi)\Box\phi+\frac{1}{2}\omega^{\prime}(\phi)h^{\mu\nu}{\cal D}_{\mu}\phi{\cal D}_{\nu}\phi-\gamma\omega^{\prime}(\phi)A^{\mu}\phi{\cal D}_{\mu}\phi+\nonumber\\
&&\gamma\omega(\phi)\nabla_{\mu}A^{\mu}-\gamma^2\omega(\phi)A^{\mu}A_{\mu}\phi+V^{\prime}(\phi)=0,\label{wm4}\\
&& \nabla_{\mu}F^{\mu\nu}= J^{\nu}-\gamma\omega(\phi)h^{\mu\nu}\phi{\cal D}_{\mu}\phi.\label{wm5}
\end{eqnarray}
It is important to note here that the Weyl scalar field couples to matter. This geometrical coupling motivates the idea that dark matter and dark energy may interact each other in this framework. 
In the next section we will propose a model for a cosmological  viscous dark fluid which can be described by the scalar field $\phi$.

\section{A viscous dark fluid model}

Now we are interested in formulate a cosmological model for the present accelerating expansion epoch, with a viscous dark fluid in the formalism previously explained. The main idea is to model the viscous fluid with the scalar field $\phi$ of geometrical origin. To implement the cosmological principle we use the gauge election in the equation \eqref{yq1} : $\sigma_{,\mu}=\gamma A_{\mu}$. Clearly, under this choice $\overset{\smile}{A}_{\mu}=0$ and thus the electromagnetic part obeys $\overset{\smile}{F}_{\mu\nu}=0$. Hence, the action \eqref{yq3} reduces in this gauge to
\begin{equation}\label{df0}
{\cal S}=\int d^{4}x\,\sqrt{-h}\left[\frac{R}{16\pi G}+\frac{1}{2}\omega(\phi)h^{\mu\nu}\phi_{,\mu}\phi_{,\nu}-V(\phi)\right].
\end{equation}
The field equations derived from this action are
\begin{eqnarray}
&& G_{\mu\nu}=8\pi G \left[\omega(\phi)\phi_{,\mu}\phi_{,\nu}-\frac{1}{2}h_{\mu\nu}\left(\omega(\phi)h^{\alpha\beta}\phi_{,\alpha}\phi_{,\beta})\right.\right.\nonumber\\
\label{df1}
&& \left.\left.-2V(\phi)\right)\right],\\
\label{df2}
&& \omega(\phi)\Box\phi+\frac{1}{2}\omega^{\prime}(\phi)h^{\mu\nu}\phi_{,\mu}\phi_{,\nu}+V^{\prime}(\phi)=0,
\end{eqnarray}
where the prime denotes derivative with respect $\phi$. 
Now, considering small deviations of the kinetic term from the canonical kinetic energy in the action \eqref{df0}, we can  assume that $\omega(\phi)$ has the form
\begin{equation}\label{df3}
\omega(\phi)=1+\epsilon \zeta(\phi),
\end{equation}
with $\epsilon \ll 1$ being a dimensionless parameter. Thus, the field equations \eqref{df1} and \eqref{df2} become
\begin{eqnarray}
&& G_{\mu\nu}=8\pi G\left[\phi_{,\mu}\phi_{,\nu}-\frac{1}{2}h_{\mu\nu}(\phi_{,\alpha}\phi^{,\alpha}-2V(\phi))+\right.\nonumber\\
\label{df4}
&&\left.\epsilon\zeta(\phi)\phi_{,\mu}\phi_{,\nu}-\frac{\epsilon}{2}h_{\mu\nu}\zeta(\phi)\phi_{,\alpha}\phi^{,\alpha}\right],\\
&& \Box\phi+V^{\prime}(\phi)+\epsilon\left[\zeta(\phi)\Box\phi+\frac{1}{2}\zeta^{\prime}(\phi)h^{\mu\nu}\phi_{,\mu}\phi_{,\nu}\right]=0.\nonumber\\ 
\label{df5}
\end{eqnarray}
On the other hand, the energy-momentum tensor for a perfect fluid with scalar viscosity is given by
\begin{equation}\label{df6}
T_{\mu\nu}^{(df)}=(\rho_{df}+p_{df})U_{\mu}U_{\nu}-p_{df}h_{\mu\nu}+\eta\nabla_{\sigma}U^{\sigma}(h_{\mu\nu}-U_{\mu}U_{\nu}),
\end{equation}
where $\eta$ is the scalar viscosity coefficient, $\rho_{df}$ and $p_{df}$ are the energy density and pressure for the dark fluid, respectively. The 4-velocity field of the fluid $U^{\lambda}$ is here given for the family of comoving cosmological observers $U^{\mu}=\delta^{\mu}_{0}$, with magnitude $U^{\alpha}U_{\alpha}=1$. \\

Now, in order to construct a dark fluid model where the viscosity is described by the non-canonical part of the scalar field, we consider the energy momentum tensor for the scalar field that appears in \eqref{df4}, which is given by 
\begin{eqnarray}
T_{\mu\nu}^{(\phi)} & = &\phi_{,\mu}\phi_{,\nu}-\frac{1}{2}h_{\mu\nu}\left(\phi_{,\alpha}\phi^{,\alpha}-2V(\phi)\right)+\epsilon\zeta(\phi)\phi_{,\mu}\phi_{,\nu}\nonumber\\
&& 
-\frac{\epsilon}{2}h_{\mu\nu}\zeta(\phi)\phi_{,\alpha}\phi^{,\alpha}.\label{df7}
\end{eqnarray}
Thus, comparing the expression \eqref{df6} with \eqref{df7} we arrive to the system
\begin{eqnarray}
\label{df8}
 && (\rho_{df}+p_{df})U_{\mu}U_{\nu}-p_{df}h_{\mu\nu}=\phi_{,\mu}\phi_{,\nu}-\nonumber\\
&&\frac{1}{2}h_{\mu\nu}\left[\phi^{,\alpha}\phi_{,\alpha}-2V(\phi)\right],\\
&& \eta\nabla_{\sigma}U^{\sigma}(h_{\mu\nu}-U_{\mu}U_{\nu})=\epsilon\zeta{\phi}\left(\phi_{,\mu}\phi_{,\nu}-\frac{1}{2}h_{\mu\nu}\phi^{,\alpha}\phi_{,\alpha}\right).\nonumber\\
 \label{df9}
\end{eqnarray}
The trace of equation \eqref{df9} results in
\begin{equation}\label{df10}
\zeta(\phi)=-\frac{3\eta\nabla_{\lambda}U^{\lambda}}{\epsilon\phi_{,\alpha}\phi^{,\alpha}}.
\end{equation}
Considering the metric in the 3D-spatially flat FRW line element
\begin{equation}\label{df11}
ds^2=dt^2-a^2(t)\left(dx^2+dy^2+dz^2\right),
\end{equation}
where $a(t)$ is the cosmic scale factor and $t$ is the cosmic time, the equation \eqref{df10} for an homogeneous and isotropic field $\phi$ becomes
\begin{equation}\label{df12}
\zeta(\phi)=-\frac{9\eta H(\phi)}{\epsilon\dot{\phi}^2},
\end{equation}
with $H=\dot{a}/a$ being the Hubble parameter. 
Now, with the help of  \eqref{df4}, \eqref{df7}, \eqref{df8}, \eqref{df9} and \eqref{df11} we arrive to the Friedmann equations 
\begin{eqnarray}\label{df13}
 3H^2 &=& 8\pi G \rho_{df},\\
\label{df14}
\dot{H}+H^2 &=& -\frac{4\pi G}{3} \left(\rho_{df}+3p_{df}-9\eta H\right).
\end{eqnarray}
Similarly, the equation for the scalar field \eqref{df5} in the FRW metric \eqref{df11} acquires the form
\begin{equation}\label{df15}
\ddot{\phi}+3H\dot{\phi}+V^{\prime}(\phi)+\epsilon\left[\zeta(\phi)\left(\ddot{\phi}+3H\dot{\phi}\right)+\frac{1}{2}\zeta^{\prime}(\phi)\dot{\phi}^2\right]=0.
\end{equation}
The equations \eqref{df14} and \eqref{df15} with the help of \eqref{df6} yield
\begin{equation}\label{df16}
\dot{\rho}_{df}+3H\left(\rho_{df}+p_{df}-3\eta H\right)=0.
\end{equation}
Assuming a dark fluid characterized by an equation of state
\begin{equation}\label{cni1}
p_{df}=\omega_{df}(\rho_{df})\rho_{df}.
\end{equation}
Inspired in the equation \eqref{df14}, we assume a form for the scalar viscosity given by
\begin{equation}\label{pof1}
\eta(H,\dot{H})=\alpha H+\beta \frac{\dot{H}}{H}=(\alpha-\beta)H-\beta qH,
\end{equation}
where $\alpha$ and $\beta$ are parameters to be determined and the deceleration parameter is $q=-a\ddot{a}/\dot{a}^2$. Using \eqref{df13}, \eqref{cni1} and \eqref{pof1} in \eqref{df16} we obtain  
\begin{equation}\label{pof2}
\dot{\rho}_{df}+3H(1+\omega_{df})\rho_{df}-24\pi G\left(\alpha H+\beta\frac{\dot{H}}{H}\right)\rho_{df}=0.
\end{equation}
This equation determines the dynamics of the energy density associated to the dark fluid, which depends of its internal composition. We are now in position to look for solutions of \eqref{pof2} for different forms of the equation of state (EOS) parameters $\omega_{df}(\rho_{ef})$.

\subsection{The case of constant dark energy EOS parameter }

The simplest case we study is when $\omega_{de}=\omega_{de}^{(0)}$ is a constant. In this particular situation the equation \eqref{pof2} reads
\begin{equation}\label{pof3}
\dot{\rho}_{df}+3H(1+\omega_{df}^{(0)})\rho_{df}-24\pi G\left(\alpha H+\beta\frac{\dot{H}}{H}\right)\rho_{df}=0.
\end{equation}
Solving \eqref{pof3} we obtain
\begin{equation}\label{pof4}
\rho_{df}=\rho_{df}^{(0)}\left(\frac{a_0}{a}\right)^{n_1}\left(\frac{H}{H_0}\right)^{n_2},
\end{equation}
where $n_1=3(1+\omega_{df}^{(0)})-24\pi G\alpha$ and $n_{2}=24\pi G\beta$, with $a_0=a(t_0)$ and $H_0=H(t=0)$ are the scale factor and the Hubble parameter evaluated in the present time $t_0$. Thus if follows from \eqref{df13} and \eqref{pof4} that the scale factor has the form
\begin{equation}\label{pof5}
a(t)=\left[a_0^{\frac{n_1}{2-n_2}}+\frac{n_1\lambda}{2-n_2}(t-t_0)\right]^{\frac{2-n_2}{n_1}},
\end{equation}
where $n_1=3(1+\omega_{df}^{(0)})-24\pi G \alpha$, $n_2=24\pi G \beta$, $a_0=a(t_0)$ and 
\begin{equation}\label{pof6}
\lambda=\left(\frac{8\pi G a_0^{n_1}}{3H_0^{n_2}}\rho_{df}^{(0)}\right)^{\frac{1}{2-n_2}}.
\end{equation}
The equation \eqref{pof5} gives the scale factor corresponding to the whole dark fluid. However, we have assumed that the fluid has two dark components: dark energy and dark matter. If we consider that these two componentes interact each other the equation \eqref{pof2} leads to the system
\begin{eqnarray}
&& \dot{\rho}_{de}+3H(\rho_{de}+\omega_{de}(\rho_{de})\rho_{de})-24\pi G\eta\rho_{de}= -Q,\nonumber\\
\label{pof7}\\
\label{pof8}
&& \dot{\rho}_{m}+3H\rho_{m}-24\pi G\eta\rho_{m}=Q,
\end{eqnarray}
where $Q$ is the interaction function, $\rho_{de}$ denotes the dark energy density and $\rho_m$ is the dark matter energy density. If we consider that $\rho_m(t)=r(t)\rho_{de}(t)$ the system \eqref{pof7} and \eqref{pof8} becomes
\begin{equation}\label{pof9}
\dot{\rho}_{de}+\left[3H\left(1+\frac{\omega_{de}(\rho_{de})}{1+r}\right)+\frac{\dot{r}}{1+r}-24\pi G\eta\right]\rho_{de}=0.
\end{equation}
With these considerations the EOS parameter can be written as
\begin{equation}\label{cni2}
\omega_{df}=\frac{\omega_{de}(\rho_{de})}{1+r}.
\end{equation}
Hence, the Friedmann equation \eqref{df13} reads
\begin{equation}\label{pof10}
H^2=\frac{8\pi G (1+r)}{3}\rho_{de}.
\end{equation}
Solving \eqref{pof9} for $\omega_{de}=\omega_{de}^{(0)}$ and $r=r_0$ both constants, we obtain
\begin{equation}\label{pof11}
\rho_{de}=\rho_{de}^{(0)}\left(\frac{a_0}{a}\right)^{n_3}\left(\frac{H}{H_0}\right)^{n_2},
\end{equation}
where we have used \eqref{pof1} and 
\begin{equation}\label{pof12}
n_3=3\left(1+\frac{\omega_{de}^{(0)}}{1+r}\right)-24\pi G\alpha.
\end{equation}
Now, inserting \eqref{pof11} in \eqref{pof10} we obtain for the scale factor
\begin{equation}\label{pof13}
a(t)=	\left(a_{0}^{\frac{n_3}{2-n_2}}+\frac{n_3\gamma}{2-n_2}(t-t_0)\right)^{\frac{2-n_2}{n_3}},
\end{equation}
where 
\begin{equation}\label{corrr1}
\gamma=\left(\frac{8\pi G}{3H_0^{n_2}}(1+r)\rho_{de}^{(0)}a_0^{n_3}\right)^{\frac{1}{2-n_2}}.
\end{equation}
Notice that this scale factor has the same form than \eqref{pof5}. This is because the both describe the expansion of the universe in the present time of accelerated expansion. The equation \eqref{pof5} takes into account the whole dark without regarding an interaction between the two dark sectors, whereas the equation \eqref{pof13} considers mainly information of the dark energy sector in a self-interacting viscous dark fluid.\\

On the other hand, it follows from \eqref{df8} that the pressure and energy dentiy of dark fluid satisfy
\begin{eqnarray}\label{agr2}
&& p_{df}=\omega_{de}\rho_{de}=\frac{1}{2}\dot{\phi}^2-V(\phi),\\
\label{agr3}
&& \rho_{df}=\rho_{m}+\rho_{de}=(1+r)\rho_{de}=\frac{1}{2}\dot{\phi}^2+V(\phi).
\end{eqnarray}
Therefore the scalar field $\phi$ and the potential $V$ obey respectively
\begin{eqnarray}\label{agr4}
&& \dot{\phi}^2=(1+r+\omega_{de})\rho_{de},\\
\label{agr5}
&& V=\frac{1}{2}(1+r-\omega_{de})\rho_{de}.
\end{eqnarray}
With the help of \eqref{pof11} and \eqref{pof13} the equation \eqref{agr4} has for solution
\begin{equation}\label{agr6}
\phi(t)=\phi_0+p_0\ln \left[1+\mu_0(t-t_0)\right],
\end{equation}
where
\begin{eqnarray}
p_0 &=& a_0^{n_3/2}\left(\frac{2-n_2}{n_3\gamma}\right)\left(\frac{\gamma}{H_0}\right)^{n_2/2}\,\sqrt{(1+r+\omega_{de})\rho_{de}^{(0)}}\,,\nonumber\\ 
\label{agr7} \\
\label{agr8}
\mu_0 &=& \frac{n_3\gamma}{2-n_2}a_{0}^{-\frac{n_3}{2-n_2}}\,.
\end{eqnarray}
Using \eqref{agr6} and \eqref{agr5} we obtain
\begin{equation}\label{agr9}
V(\phi)=V_{0c}\exp\left[-\frac{2}{p_0}(\phi-\phi_0)\right],
\end{equation}
where 
\begin{equation}\label{agr10}
V_{0c}=\frac{1}{2}(1+r-\omega_{de})\rho_{de}^{(0)}a_{0}^{-\frac{n_2n_3}{2-n_2}}\left(\frac{\gamma}{H_0}\right)^{n_2}.
\end{equation}
Now, employing \eqref{pof13} the present deceleration  parameter $q_0=-(1+\dot{H}_0/H_0^2)$ reads
\begin{equation}\label{pan1}
q_0=-\left(1-\frac{n_3}{2-n_2}\right).
\end{equation}
According to the Planck 2018 results $q_0=-0.5581^{+0.0273}_{-0.0267}$ \cite{Planck2018}. It follows from \eqref{pan1} that
\begin{equation}\label{pan2}
\alpha=\frac{1}{8\pi G}\left(1+\frac{\omega_{de}^{(0)}}{1+r}\right)-\frac{1}{12\pi G}(1+q_0)(1-12\pi G\beta),
\end{equation}
where we have used the definitions of $n_2$ and $n_3$. Thus, if the condition \eqref{pan2} holds, then the model is in agreement with observations for the present values of the deceleration parameter.\\

With the help of \eqref{df12}, \eqref{pof13} and \eqref{agr6} we arrive to
\begin{equation}\label{pan3}
\zeta(\phi)=-\frac{9\eta\gamma}{\epsilon p_0^2\mu_0^2}\frac{\exp\left(\frac{2}{p_0}(\phi-\phi_0)\right)}{a_{0}^{\frac{n_3}{2-n_2}}+\frac{n_3\gamma}{(2-n_2)\mu_0}\left[\exp\left(\frac{1}{p_0}(\phi-\phi_0)\right)-1\right]}.
\end{equation}
This expression corresponds to 
\begin{equation}\label{pan4}
\omega(\phi)=1-\frac{9\eta\gamma}{ p_0^2\mu_0^2}\frac{\exp\left(\frac{2}{p_0}(\phi-\phi_0)\right)}{a_{0}^{\frac{n_3}{2-n_2}}+\frac{n_3\gamma}{(2-n_2)\mu_0}\left[\exp\left(\frac{1}{p_0}(\phi-\phi_0)\right)-1\right]}.
\end{equation}
In this manner, a deviation from the canonical kinetic energy of the scalar field $\phi$ given by \eqref{pan3} allows to describe a dark fluid with constant EOS parameter from a single scalar field.

\subsection{The case of a thermodynamical dark energy EOS parameter.}

As it was done in \cite{Myrzakul, Myrzakulov}, we consider a thermodynamical EOS parameter of the form
\begin{equation}\label{cni9}
\omega_{de}(\rho_{de})=A\rho_{de}^{n-1}-1,
\end{equation}
where $n > 1$. 
Using \eqref{cni9} the equation \eqref{pof9} yields
\begin{eqnarray}
&& \dot{\rho}_{de}+\left[3H\left(1-\frac{1}{1+r}\right)+\frac{\dot{r}}{1+r}-24\pi G\eta\right]\rho_{de}\nonumber\\
&& 
 +\frac{3AH}{1+r}\rho_{de}^{n}=0.\label{pof14}
\end{eqnarray}
This equation can be written as
\begin{eqnarray}
&&\dot{Z}_{de}+(1-n)\left[3H\left(1-\frac{1}{1+r}\right)+\frac{\dot{r}}{1+r}-24\pi G\eta\right]Z_{de}\nonumber\\
&& =-\frac{3AH}{1+r},\label{pof15}
\end{eqnarray}
where $Z_{de}=\rho_{de}^{-(n-1)}$. Solving \eqref{pof15} we obtain
\begin{eqnarray}
&& Z_{de}(t) = 3A(n-1)\frac{a^{(n-1)(3-24\pi G\alpha)}}{H^{24\pi G\beta (n-1)}}(1+r)^{n-1}\times\nonumber\\
&&
\exp\left(-3(n-1)\int_{t_i}^{t}dt\,\frac{H}{1+r}\right)\left[\int_{t_i}^{t}dt(1+r)^{-n}\times\right.\nonumber\\
\label{pof16} 
&& a^{-(n-1)(3-24\pi G\alpha)}H^{24\pi G\beta (n-1)+1}\times\nonumber\\
&&
\left.\exp\left(3(n-1)\int_{t_i}^{t}dt\,\frac{H}{1+r}\right)\right],
\end{eqnarray}
where $t_i$ is the time when the accelerated expansion started. Hence, the density of dark energy reads
\begin{eqnarray}
\rho_{de}(t) &=& \frac{[3A(n-1)]^{-\frac{1}{n-1}}}{1+r}\frac{H^{n_2}}{a^{3-24\pi G\alpha}}\times\nonumber\\
&&
\exp\left(3\int_{t_i}^{t}dt\,\frac{H}{1+r}\right)\left[\int_{t_i}^{t}\,(1+r)^{-n}\times\right.\nonumber\\
\label{agr1}
&&  a^{-(n-1)(3-24\pi G\alpha)}H^{1+n_2(n-1)}\times\nonumber\\
&&
\left.\exp\left(3(n-1)\int_{t_i}^{t}dt\, \frac{H}{1+r}\right)\right]^{-\frac{1}{n-1}}.
\end{eqnarray}
However, according to obervational data the present values of the densities of dark energy and dark matter obey $\rho_{de0}/\rho_{dm0}\sim{\cal O}(1)$. The question about ``why now" constitutes the cosmological coincidence problem. Moreover, it follows from the standard model that the equality $\rho_{de}=\rho_{dm}$ took place ``recently", at a redshift $z\approx 0.55$ \cite{Zim}. Thus we can assume a slowly enough time variation  of the $r$ parameter as to be practically considered as constant. In this manner, for constant $r$ the expression \eqref{pof16} becomes
\begin{equation}\label{pof17}
Z_{de}(t)=\frac{3A(n-1)}{1+r}\frac{a^{n_A}}{H^{n_2}}\int_{t_i}^{t} dt\, a^{-n_A}H^{1+n_B},
\end{equation}
where $n_B=n_2(n-1)$ and $n_A=(n-1)[3(1-1/(1+r))-24\pi G\alpha]$. Considering $\beta\ll 1$ in the viscosity \eqref{pof1}, the equation \eqref{pof17} leads to
\begin{equation}\label{pof18}
\rho_{de}(a)=\rho_{de}^{(0)}\left[\left(\frac{a}{a_i}\right)^{n_A}-1\right]^{-\frac{1}{n-1}},
\end{equation}
where $\rho_{de}^{(0)}=[3A(n-1)/(n_A(1+r))]^{-1/(n-1)}$. With the help of \eqref{pof18} the equation \eqref{pof10} for $a>a_i$ gives a scale factor of the form
\begin{eqnarray}
a(t)&=&a_{i}\left[1+\frac{n_A}{2(n-1)}\sqrt{\frac{8\pi G}{3}(1+r)\rho_{de}^{(0)}}\,(t-t_i)\right]^{\frac{2(n-1)}{n_A}}\nonumber\\
\label{pof19}
&=& a_{i}\left[1+\lambda_0(t-t_i)\right]^{\frac{2(n-1)}{n_A}}.
\end{eqnarray} 
It is not difficult to see that when $\lambda_0 (t-t_i)\gg 1$ the dark energy density as time function reads
\begin{equation}\label{pof20}
\rho_{de}(t)=\frac{\rho_{de}^{(0)}}{\lambda_0^2}\frac{1}{(t-t_i)^2},
\end{equation}
where $\lambda_0=[n_A/2(n-1)]\sqrt{(8\pi G (1+r)\rho_{de}^{(0)})/3}$. Now, according to \eqref{agr4} and \eqref{cni9} we arrive to
\begin{equation}\label{pof21}
\dot{\phi}=\sqrt{r\rho_{de}+A\rho_{de}^n}. 
\end{equation}
For $t\gg t_i$ the substitution of \eqref{pof20} in \eqref{pof21} results in the expression
\begin{equation}\label{pof22}
\dot{\phi}=\frac{\sqrt{r\rho_{de}^{(0)}\lambda_0^{-2}}}{t}\,\sqrt{1+\frac{A}{r}(\rho_{de}^{(0)})^{n-1}\lambda_0^{2-2n}\frac{1}{t^{2(n-1)}}}.
\end{equation}
For $n$ large enough we can use the approximation formula $(1+x)^n\simeq 1+nx$ in \eqref{pof22} and its solution is given by
\begin{equation}\label{pof23}
\phi(t)=\phi_i+\sqrt{r\rho_{de}^{(0)}\lambda_0^{-2}}\,\ln\left(\frac{t}{t_i}\right).
\end{equation}
The potential resulting from \eqref{agr4} and \eqref{cni9} reads
\begin{equation}\label{pof24}
V=\frac{1}{2}\left[(2+r)\rho_{de}-A\rho_{de}^n\right].
\end{equation}
By means of \eqref{pof20} the potential \eqref{pof24} can be written as
\begin{equation}\label{pof25}
V(t)=\frac{1}{2}\left[\frac{(2+r)\rho_{de}^{(0)}\lambda_{0}^{-2}}{(t-t_i)^2}-\frac{A(\rho_{de}^{(0)})^n\lambda_0^{-2n}}{(t-t_i)^{2n}}\right].
\end{equation}
Thus for $t\gg t_i$ the potential $V$ as function of the field $\phi$ is given by
\begin{eqnarray}
V(\phi)&=&\frac{(2+r)\rho_{de}^{(0)}}{2\lambda_0^2 t_i^2}\exp\left[-\frac{2\lambda_0}{\sqrt{r\rho_{de}^{(0)}}}(\phi-\phi_i)\right]-\nonumber\\
\label{pof26}
&&\frac{A(\rho_{de}^{(0)})^n}{2\lambda_0^{2n}t_{i}^{2n}}\exp\left[-\frac{2n\lambda_0}{\sqrt{\rho_{de}^{(0)}}}(\phi-\phi_i)\right].
\end{eqnarray}
On the other hand, the deceleration parameter corresponding to \eqref{pof19} is given by
\begin{equation}\label{pof27}
q_0=-\left(1-\frac{n_A}{2(n-1)}\right).
\end{equation}
By means of the definition of $n_A$ the expression \eqref{pof27} reduces to 
\begin{equation}\label{pof28}
\alpha=\frac{1}{8\pi G}\left(1-\frac{1}{1+r}\right)-\frac{1}{12\pi G}(1+q_0).
\end{equation}
Thus for $\alpha$ given by \eqref{pof28} the deceleration parameter in this model is in agreement with Planck 2018 results. Finally using \eqref{df12}, \eqref{pof19} and \eqref{pof23} we obtain
\begin{equation}\label{pof29}
\zeta(\phi)=-\zeta_0\left[\frac{\exp\left(\frac{2\lambda_0}{\sqrt{r\rho_{de}^{(0)}}}(\phi-\phi_i)\right)}{1+t_i\lambda_0\left(\exp\left(\frac{\lambda_0}{\sqrt{r\rho_{de}^{(0)}}}(\phi-\phi_i)\right)-1\right)}\right],
\end{equation}
where 
\begin{equation}\label{coes}
\zeta_0=\frac{18\eta(n-1)\lambda_0^3 t_i^2}{\epsilon r\rho_{de}^{(0)}n_A}.
\end{equation}
Therefore 
\begin{equation}\label{pof30}
\omega(\phi)=1-\epsilon\zeta_0\left[\frac{\exp\left(\frac{2\lambda_0}{\sqrt{r\rho_{de}^{(0)}}}(\phi-\phi_i)\right)}{1+t_i\lambda_0\left(\exp\left(\frac{\lambda_0}{\sqrt{r\rho_{de}^{(0)}}}(\phi-\phi_i)\right)-1\right)}\right].
\end{equation}
Thus, an interacting dark fluid scenario with a thermodynamical dark energy EOS parameter can be derived from a geometrical scalar-tensor theory of gravtity when we use \eqref{pof30} in the action \eqref{df0}.

\section{Final Remarks}

In this letter we have investigated the possibility that viscous cosmological dark fluid models can be described in a unified manner by a single scalar field, in the context of an invariant geometrical scalar-tensor theory of gravity.  In this new version of scalar-tensor theories the main feature is that both the metric tensor and the scalar field have geometrical origin, in virtue that they appear in the affine conection of the background geometry, which in this case results to be the Weyl-Integrable one. As it is shown in the action \eqref{df0} in the Riemann frame the scalar field has a non-canonical kinetic energy. When $\omega(\phi)=1$ the kinetic energy is canonical.\\

We found that for the cases of constant and themodynamic equation of state parameters the unified description is possible. According to the expressions \eqref{df3}, \eqref{df8} and \eqref{df9} the scalar viscosity of the fluid appears when the kinetic energy of the scalar field deviates from the canonicity. Thus a viscous dark fluid with a constant EOS parameter can be described by a non-canonicity parameter given by \eqref{pan4} and a scalar potential \eqref{agr9}. In the case of the thermodynamic EOS parameter \eqref{cni9}, the scalar description of the fluid is determined by the expressions \eqref{pof30} and \eqref{pof26}. Something interesting is that the thermodynamical case reduces to the constant EOS when $\rho_{de}=1/(A-\omega_{de}^{(0)})$. In particular the $\Lambda$-CDM model is recovered when $A= 8\pi/(\Lambda M_p^2)-\omega_{de}^{(0)}$. \\

As it is known, in different alternative theories of gravity like for example $F(R)$ theories, dark energy viscous models can be described. Moreover, scenarios of unification of the present accelerated expansion period with the early-time inflationary epoch were also studied \cite{rea15,rea37}. In our case,  as it was shown in \cite{rea32,rea33}, in the context of geometrical scalar tensor theories some inflationary scenarios including Higgs inflation can be modeled by means of the same  scalar field formalism we use in here. In this manner, the fact to unify by a scalar field a viscous dark fluid opens the possibility to unify the present period of accelerated expansion with Higgs inflation, in the context of geometrical invariant scalar-tensor theories of gravity. It is important to emphasize that these theories are free of the frames controversy we mention in the introduction.

\section*{Acknowledgements}

\noindent  J.E.Madriz-Aguilar  acknowledges CONACYT
M\'exico and Centro Universitario de Ciencias Exactas e Ingenierias and Centro Universitario de los Valles of Universidad de Guadalajara for financial support. V. A. Gil-Ocaranza and J. Zamarripa acknowledge Centro Universitario de los Valles of Universidad de Guadalajara for financial support. M. Montes acknowledges Centro Universitario de Ciencias Exactas e Ingenierias of Universidad de Guadalajara for financial support.

\end{document}